\documentclass{vgtc}                          

\graphicspath{{figures/}{pictures/}{images/}{./}} 

\usepackage{times}                     

\usepackage{tabu}                      
\usepackage{booktabs}                  
\usepackage{lipsum}                    
\usepackage{mwe}                       
\usepackage{pdfpages}   
\usepackage{mathptmx}                  

\onlineid{0}

\vgtccategory{Research}

\vgtcinsertpkg

\title{Bringing Journaling Data to Life: \\
Designing Data Characters 
for the Emotional Self}

\author{
Diego Abarcar Calugay\thanks{e-mail: diegoac@terpmail.umd.edu}
\and Isabella Amador\thanks{e-mail: iamador1@terpmail.umd.edu}
\and Keke Wu\thanks{e-mail: kekewu@umd.edu}}
\affiliation{\vspace{-0.5em}\scriptsize University of Maryland, College Park}

\usepackage{graphicx} 

\teaser{
  \centering
  \includegraphics[width=\linewidth]{teaser_pictorial_interface_wider.pdf}
  \caption{Creating Data Characters from journaling data. Users customize facial features and body posture to construct personalized, human-like representations of affective experiences.
}
  \label{fig:teaser}
}

\abstract{Journaling is a common practice for emotional expression, reflection, and processing. However, as entries accumulate, it can become difficult to interpret and compare their affective content, especially since traditional text-based analyses and visualizations often struggle to convey affective nuance. We introduce Data Characters, a visualization approach that represents affective content in journaling through human-like characters. Using a customizable Data Character as a design probe, we investigate the potential of character-based representations for conveying affective experiences and explore what visual encodings emerge through customization. Preliminary walkthroughs with two participants demonstrate the intuitiveness and feasibility of the approach. This work contributes an exploratory approach to studying how affective experiences can be visually represented and encoded through anthropomorphic forms.}

\keywords{Visual representation design, Personal visualization, Personal visual analytics.}

\begin{document}
\setlength{\parskip}{0pt}

\firstsection{Introduction}

\maketitle

Journaling, the practice of documenting life experiences and personal reflections through writing, is widely used to process, regulate, and reflect on emotional experiences \cite{journaling}. However, as entries accumulate, identifying affective patterns, comparing experiences, and understanding emotional changes can become difficult.

Data visualization offers a promising medium to support such interpretation. However, existing text visualizations often emphasize word- or sentence-level features, such as frequency, overlooking context and emotional nuance \cite{QualVis}. More structured approaches, such as thematic coding, reduce text to discrete categories (e.g., ``happy'' or ``sad''), potentially oversimplifying complex affective experiences \cite{QualVis}. This motivates new visualizations that more directly represent affective content in personal data. 

Inspired by nonverbal communication \cite{NonverbalEmotionalExpression} and anthropomorphic thinking \cite{AnthroPsych}, we introduce \textit{Data Characters}, a visualization approach that represents affective content in journaling through human-like characters' facial expressions and body postures~\cite{DataChar}. Using a customizable Data Character as a design probe, we allow people to construct their own representations of affective experiences. Through the customization process, we explore the potential of character-based representations and investigate how character features may function as affective visual encodings.



\section{Design Rationale}

Data Characters were motivated by a simple observation---people naturally communicate emotions through facial expressions and body posture. Rather than relying on abstract visual forms, we sought a representation that used these familiar cues. This led us to \textit{anthropomorphism}: using human-like qualities to express affect through a data representation. 

Anthropomorphism has appeared in visualization through anthropographics~\cite{AnthroMorais-DesignSpace}, designs that humanize data to elicit empathetic responses. Despite their contested effectiveness, 
prior work suggests that human-like features may support emotional expression and interpretation. Objects resembling human postures can be perceived as emotionally expressive \cite{FlexibleDisplay}, while anthropomorphic faces and voices have been found to support understanding of complex instructions \cite{AnthroPsych}. Together, these findings suggest that anthropomorphism may offer a promising way to represent affective content.

Building on these ideas, we developed Data Characters: a visualization approach that represents affective experiences through full-body, human-like characters. Rather than prescribing how emotions should map to character features, we make these features customizable and ask: \\

\smallskip
\noindent\hangindent=2.7em\hangafter=1
\textbf{RQ1.} How can Data Characters represent affective experiences?

\noindent\hangindent=2.7em\hangafter=1
\textbf{RQ2.} What visual encodings emerge through customization?
\smallskip

\section{Data Characters as a Design Probe}

\begin{figure}[t]
    \centering
    \includegraphics[width=\columnwidth]{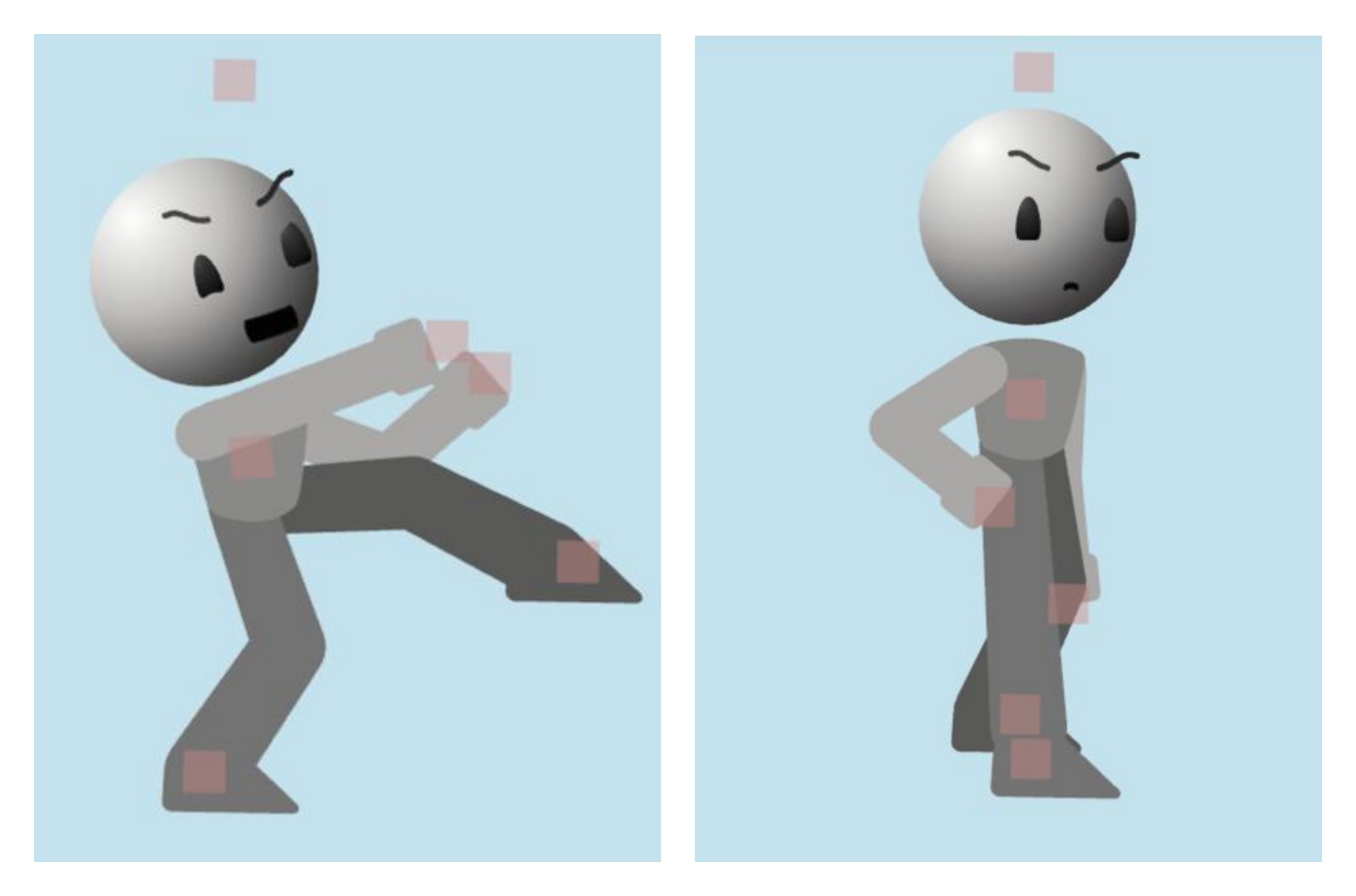}
    \caption{User-generated visualizations of ``anger'' illustrate personalized representations of the same emotion.}
    \label{fig:pad}
\end{figure}

To investigate our research questions, we implemented a customizable Data Character (Fig. \ref{fig:teaser}) as a design probe \cite{DesignProbe}. Instead of prescribing how affective content maps to visual features, the interface allows users to represent affective experiences by manipulating the character's facial features and body posture. 

 
Our interface draws from prior work that describes emotions as continuous and multidimensional states: the PAD Emotional State Model \cite{PAD} and the Affective Slider \cite{AffectiveSlider}. Inspired by these instances of affective representation, our system allows users to capture a range of nuanced emotions. Users adjust the character's facial features through sliders that interpolate between expressions (e.g., frown to smile). Customizable features include mouth curvature and aperture, eye closure, and eyebrow angle, which are important for recognizing emotion expressions \cite{FacialE}.

Users can then customize the character's posture through drag-and-drop controls, manipulating limb openness and flexion, as well as the orientation and lean of the head and body. These features capture postural cues important for affective expression \cite{BodyE}. The customizable Data Character was designed and animated in Rive and implemented as a web application using HTML, CSS, and JavaScript. Finalized characters can be exported as JSON containing customization values for subsequent analysis.\\

\noindent\textbf{Preliminary Walkthroughs.} We conducted preliminary walkthroughs with two participants to assess the intuitiveness and feasibility of the interface. Participants were asked to use the Data Character to represent ``anger'' (Fig. \ref{fig:pad}). Both found the interface easy to use and felt that the customization options allowed them to express emotion in personalized and nuanced ways.

\section{Implications}
Data Characters highlight new opportunities for representing affective content in personal text data. Journal entries contain rich affective information, yet many existing qualitative visualizations emphasize lexical, thematic, or statistical properties of text. We suggest that affective experiences themselves can be visualized, providing an alternative way to interact with journal data. Moreover, Data Characters explore how anthropomorphic representations can leverage familiar cues for emotional expression. Rather than relying  on abstract visuals, character-based designs offer a familiar and expressive way to represent affect \cite{AnthroPsych}. Finally, Data Characters may offer new ways to communicate affective experiences, supporting the interpretation of emotional patterns across time and between individuals, including in collaborative and therapeutic settings. 

Our work opens a design space for direct affective representation, offering designers a way to engage with emotions as visualizable content and motivating further research into how affect can be visually encoded.

\section{Limitations \& Future Work}

Our preliminary walkthroughs involved two participants representing a single emotion, providing initial evidence of the feasibility and intuitiveness of Data Characters rather than a systematic evaluation. We aim to expand this work with a larger design-probe study examining a range of affective experiences and the encoding strategies that emerge. Our current approach raises several questions about the scope and boundaries of Data Characters as a representational approach.

First, Data Characters convey affect with nonverbal expressions that may not be universal \cite{NonverbalEmotionalExpression}, limiting the generalizability of these representations. Future work should examine how user-generated representations are interpreted across individuals and cultures.

Second, anthropomorphizing emotions may not be intuitive or expressive for everyone. Future work could investigate other techniques (e.g., visual metaphors) to better understand different strategies for affective representation \cite{affectivevis}.

Third, our system focuses primarily on affect but does not afford representation of the broader context in which emotions occur. Journal entries often describe interactions, relationships, and events that shape affective experiences. Future work could explore how affective and contextual information might be represented together.


\section{Conclusion}
We introduced \textit{Data Characters}, a novel visualization approach for representing affective content in journaling. Using a customizable Data Character as a design probe, we examined the potential of character-based representations and how facial and bodily features may serve as affective visual encodings. This work contributes an exploratory approach to studying how affective experiences can be visually represented and highlights anthropomorphism as a promising approach to visualizing affective and experiential data.


\bibliographystyle{abbrv-doi-hyperref}
\bibliography{VISREF}


\includepdf[
  pages=1,
  noautoscale=true,
  width=\paperwidth,
  height=\paperheight,
  pagecommand={}
]{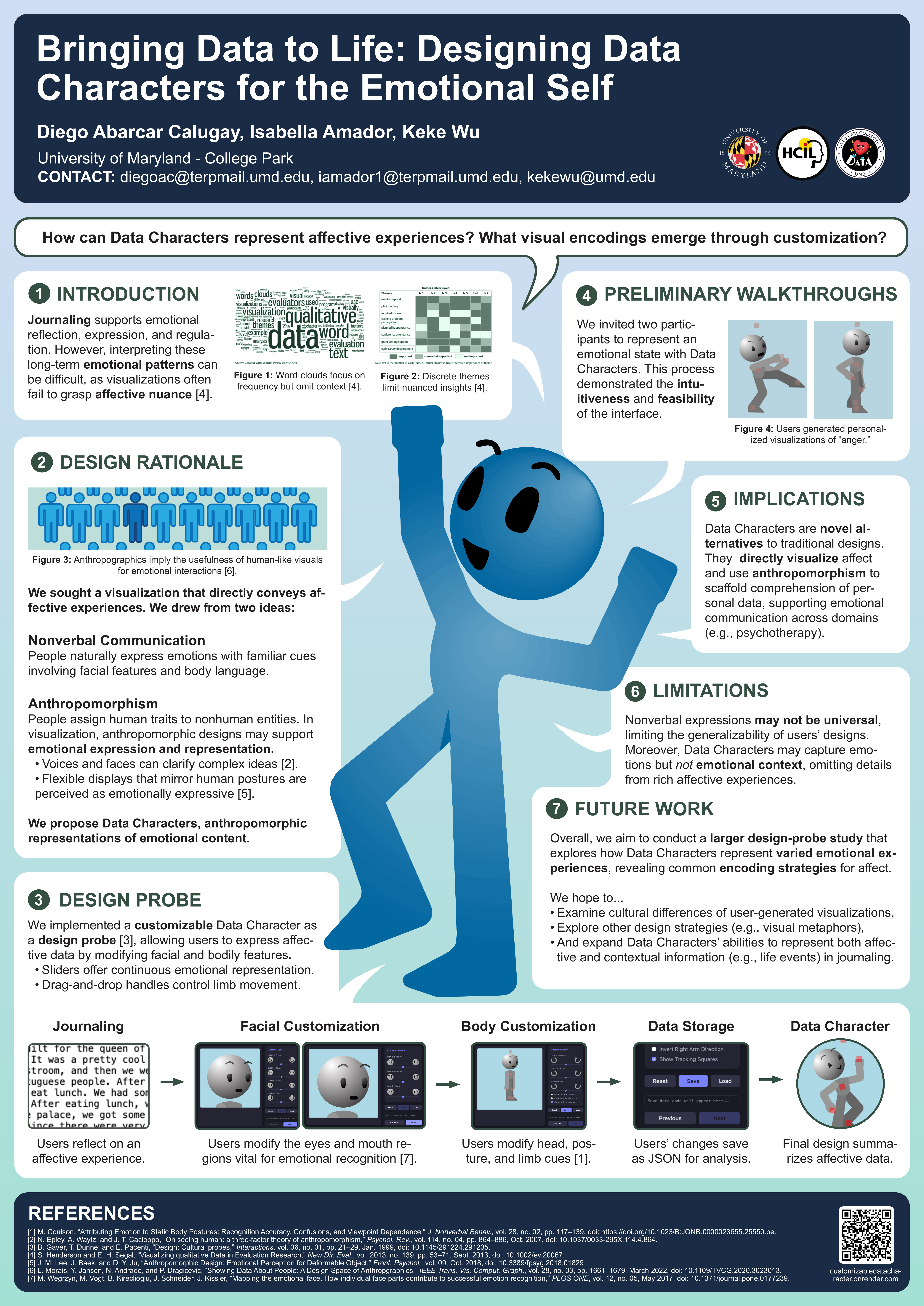}

\end{document}